**Biologically Inspired Networking and Sensing:** Algorithms and Architectures

# The Dendritic Cell Algorithm for Intrusion Detection

**Feng Gu**
*University of Nottingham, UK*

**Julie Greensmith**
*University of Nottingham, UK*

**Uwe Aicklein**
*University of Nottingham, UK*

**ABSTRACT**

As one of the solutions to intrusion detection problems, Artificial Immune Systems (AIS) have shown their advantages. Unlike genetic algorithms, there is no one archetypal AIS, instead there are four major paradigms. Among them, the Dendritic Cell Algorithm (DCA) has produced promising results in various applications. The aim of this chapter is to demonstrate the potential for the DCA as a suitable candidate for intrusion detection problems. We review some of the commonly used AIS paradigms for intrusion detection problems and demonstrate the advantages of one particular algorithm, the DCA. In order to clearly describe the algorithm, the background to its development and a formal definition are given. In addition, improvements to the original DCA are presented and their implications are discussed, including previous work done on an online analysis component with segmentation and ongoing work on automated data preprocessing. Based on preliminary results, both improvements appear to be promising for online anomaly-based intrusion detection.

Key Terms: Artificial Immune Systems, Dendritic Cell Algorithm, Danger Theory, Intrusion Detection, Botnet, Signal, Antigen, Deterministic DCA, Negative Selection, Segmentation, Correlation Coefficient, Information Gain, Principal Component Analysis.

**INTRODUCTION**

Artificial Immune Systems (AIS) (de Castro and Timmis, 2003) are computer systems inspired by both theoretical immunology and observed immune functions, principles and models, which are applied to real world problems. The human immune system, from which AIS draw inspiration, is evolved to protect the host from a wealth of invading microorganisms. AIS are developed to provide the similar defensive properties within a computing context. Initially AIS were based on simple models of the human immune system. As noted by Stibor *et al*. (2005), "first generation algorithms", including negative selection and clonal selection do not produce the same high quality performance as the human immune system. These algorithms, negative selection in particular, are prone to problems with scaling and the generation of excessive false alarms when used to solve problems such as network based intrusion detection. Recent AIS use more rigourous and up-to-date immunology and are developed in collaboration with modellers and immunologists. The resulting algorithms are believed to encapsulate the desirable properties of immune systems including robustness, error tolerance, and self-organisation (de Castro and Timmis, 2003).

One such "second generation" AIS is the Dendritic Cell Algorithm (DCA) (Greensmith, 2007), inspired by the function of the dendritic cells (DCs) of the innate immune system. It incorporates the principles of a key novel theory in immunology, termed the "danger theory" (Matzinger, 2002). This theory suggests that DCs are responsible for the initial detection of invading microorganisms, in addition to the induction of various immune responses against such invaders. An abstract model of natural DC behaviour is used as the foundation of the developed algorithm. The DCA has been successfully applied to numerous computer security related, more specific, intrusion detection problems, including port scan detection (Greensmith, 2007), botnet detection (Al-Hammadi et al., 2008) and a classifier for robot security (Oates et al., 2007). According to the results, the DCA has shown not only good performance in terms of detection rate, but also the ability to reduce the rate of false alarms in comparison to other systems, including Self Organising Maps (SOM) (Greensmith et al., 2008).

The main aim of this chapter is to demonstrate the reason for why the DCA is a suitable candidate for intrusion detection problems. In order to clearly describe the algorithm, the background and a formal definition are given. In addition, improvements to the original DCA are presented and their implications are discussed. The chapter is organised as follows: firstly, background information about a series of well known AIS algorithms and intrusion detection are described in section 2; secondly, several population AIS approaches for intrusion detection are introduced in section 3; thirdly, the algorithm details and formal definition of the DCA are demonstrated in section 4; fourthly, issues with the current DCA and potential solutions are discussed in section 5; finally, a summary of the work and some future directions are given in section 6.

**BACKGROUND**

**Intrusion Detection**

Intrusion detection involves the detection of any disallowed activities in networked computer systems. Based on deployment, intrusion detection systems can be grouped into either host-based or network-based. Host-based intrusion detection refers to the systems that monitor and collect data from the host machine. Data can be log files that include system information, such as CPU usage, memory usage, incoming/outgoing network traffics, and information of processes that are running on the host. Conversely, network-based intrusion detection refers to the systems that monitor and collect network traffic data among multiple hosts that are required for protection. Each host is a source of monitoring and collecting data, termed 'sensor'. Generally, network traffic is represented by network packets, which contain information of communications between the sources and destinations, such as IP address, port, service etc. Nowadays, most intrusion detection systems are hybrids of host-based and network-based deployments (Bejtlich, 2005).

Another way of categorising intrusion detection systems is based on detection methods, namely signature-based detection and anomaly-based detection (NIST, 2001). Signature-based detection, also known as misuse detection, distinguishes an intruder by comparing patterns or signatures of the intruder with previously known intrusions. As soon as any matches occur, an alarm is triggered. Whereas, anomaly-based detection involves discriminating between normal and anomalous data, based on the knowledge of the normal data. Thus, firstly the system needs to generate profiles of normality by either training or statistical analysis. During detection, anything that deviates from the normal profile is classified as anomalous, and an alarm is launched. Both signature based and anomaly based detection have different strengths and weaknesses. Signature-based detection produces high detection rate and low rate of false alarms, but it can only recognise the intruders or attacks are previously seen. Anomaly-based detection is capable of novel intruders and attacks that have not been seen before, but it encounter the problem of generating relatively high rate of false alarms. This problem stems from the way in which normal profiles are generated, as in most cases the collected data are just a small sample and thus not representative of the whole problem. Current techniques are often unable to cope with the dynamic changes of normal profile in real world problems and complex systems, such as large computer networks in which massive amount of nodes and uncertainties are presented.

**AIS Algorithms**

AIS researchers believe that AIS are intended to perform similar functions to the natural immune, such as defence and maintenance of the host, in a computational context. Unlike genetic algorithms, there is no one archetypal AIS, instead there are four major AIS paradigms, including the Negative Selection Algorithm (NSA) (Hofmeyr &Forrest, 1999), the Clonal Selection Algorithm (CSA) (de Castro &Von Zuben, 2000), the algorithms based on idiotypic networks models (Hart & Timmis, 2008), and the Dendritic Cell Algorithm (DCA) (Greensmith, 2007). These algorithms map the defence function of the immune system, in which certain immune entities and their functions or behaviour are included. Commonly seen immune entities are antigens which are protein particles recognised by the immune system, immune cells that perform certain immune functions individually or collaboratively, and antibodies that are detectors capable of recognising and binding to antigens. These algorithms generally involve generating effective detectors that can recognise the invading intruders and induce reactions against potential threats.

The NSA are inspired by the behaviour of a population of immune cells, named "T-cells", which belong to the adaptive immune system. In order to become functional, natural T-cells undergo "negative selection" for maturation in the thymus. First of all, the immune system generates a population of naive T cells with random specificity. Any naive T-cells that are reactive to self components are then removed from the population during this process. Remaining T-cells should only react to non-self substances, which according self/non-self theory (Coico et al., 2003) are threats to the host. From an algorithmic point view, the NSA has the following steps:
1. The system initialises certain amount of random detectors, named "naive detectors";
2. The system generates a self set from the training data that only contains normal data instances;
3. The naive detectors are compared with the self set, to produce a population of detectors that only react to intruders, termed "mature detectors";
4. The data instances in the testing data are compared with each mature detector, and if any detector reacts to the incoming data instance, an alarm is triggered.

Comparisons in training and testing (detection) are accomplished by certain matching functions, to assess the affinity of two compared candidates, where an activation threshold is applied.

The CSA are based on Burnett's theory of clonal selection and immune memory (Coico et al., 2003). This includes several adaptive and learning processes. It involves another population of immune cells, named "B-cells", which produce antibodies capable of detecting diverse and numerous patterns of invading threats. Immune memory is composed of "memory cells" that are able to remember the previously seen threats. The procedures of the CSA are as follows:

1. The system initialises certain amount of random solutions to a given problem;
2. These solutions are then being exposed to training;
3. In each generation, the best solutions are selected based on a affinity measure, to reproduce with multiple clones;
4. Current solutions then undergo a mutation process with high frequency, which is proportional to the affinity measure;
5. Optimal solutions are selected in the current population, to form memory cells;
6. The steps above keep repeating until a termination point is reached.

In brief, the system also uses pattern recognition mechanism for information processing, to make decisions of selecting the optimal solutions. The probability of a solution being replaced is proportional to its goodness. The better a solution is, the more clones are reproduced. As a result, majority of the population are optimal solutions.

Idiotypic network based algorithms are derived from the immune network theory proposed by Jerne (Jerne, 1974). It suggests that the immune system can be seen as a network in which immune entities interact with each other even when antigens are absent. The interactions can be initialised not only between antigens and antibodies, but also between antibodies and antibodies. This induces either stimulating or suppressive immune responses. They result a series of immunological behaviours, including tolerance and memory emergence. There are three major factors that affect the stimulation level of B-cells (Timmis et al., 2008), which are the contribution of the antigen binding, the contribution of neighbouring B-cells, and the suppression of neighbouring B-cells. As the stimulation level of a B-cells increases, the amount of clones it produces increases accordingly. At the population level, this results a diverse set of B-cells. In addition, three mutation mechanisms are introduced, including crossover, inverse and point mutation. In principle, the idiotypic network based algorithms are similar to the CSA, apart from the interactions between solutions in the repertoire, which may result a higher convergence rate.

**APPLICATIONS OF AIS TO INTRUSION DETECTION**

Since AIS are designed though mimicking certain properties of the natural immune system, especially the detection mechanism for intruders, obvious applications of AIS can be intrusion detection problems (Kim et al., 2007). However, not all AIS paradigms are applicable, majority of the applications involve the NSA, the idiotypic network based algorithms and the DCA. In this sections, a number of noticeable applications are described.

**The NSA-based Approaches**

Initial development of an NSA based intrusion detection system was by Hofmeyr et al. (1999). The system aimed to solve problems that involve detecting malicious connections between one computer in the Local Area Network (LAN) and one external computer. Connections are represented as 49-bit binary strings that include the source IP, destination IP, source port, destination port and service type etc. Detection is performed through string matching between connection string $s$ and detector string $d$, using r-contiguous bits rule (if $s$ and $d$ have the same symbols in at least $r$ contiguous bits positions). The value $r$ is a threshold that determines the specificity of a detector. The system has a training phase where negative selection mechanism is used. Later on, a real-valued NSA was proposed by González & Dasgupta (2003), in which data representation is changed from binary strings to real-valued numbers. The use of such data representation was intended to speed up detector generation process.

As pointed out in (Stibor et al., 2005), the NSA (no matter binary string version or real-valued version) have numerous problems: firstly they cannot avoid the curse of dimensionality, and not applicable to data with high dimensionality; secondly, the detector generation can result holes within the shape-space, so it is difficult to cover just non-self, and it either generates the population of detectors covering both non-self and unseen self (over-fit) or only covering partial non-self (under-fit); thirdly, the NSA still take extensive time period to generate the adequately complete set of detectors. Therefore, negative selection is insufficient and not suitable for anomaly detection.

**The Idiotypic Network Approach**

Ostaszewski et al (2008) proposed an adaptive and dynamic intrusion detection system based on idiotypic network. This system was designed mainly for the detection of a special type of attacks, namely "denial of service" (DOS). The DAPRA 1999 dataset (MIT Lincoln Lab, 1998 ) was used to test the system. As the aim was to provide comprehensive information of network behaviour, rather than raise alarms whenever inappropriate activities are identified, the evaluation of system performance also considered the amount of information generated during usual events.

The proposed system provides a means of gathering information about incoming, or proceeding attack from the very beginning, to take efficient countermeasures against threats. According to the experiments, monitoring activities of the idiotypic network helped to identify DOS attacks in the tested data, and different kinds of attacks showed their own particular impacts on the system. In addition, extra information can be extracted from the idiotypic network to facilitate the identification of anomalies.

**The Danger Theory Approach**

The DCA was developed to overcome the problems shown with the NSA. It was initially proposed and developed by Greensmith et al. (2005) where it was applied to a basic and standard machine learning dataset, the UCI Wisconsin Breast Cancer dataset (UCI). For this simple dataset, a classification accuracy of 99% was produced. The DCA was then applied to the ping scan detection (Greensmith et al., 2006) in computer security. The results showed that the algorithm could achieve 100% classification accuracy when appropriate thresholds are used. The DCA was later on applied to SYN scan detection (Greensmith & Aickelin, 2007) where the collected dataset consists of over five million data instances. The detection scenario was that the SYN scan was launched from a victim machine, where the DCA is used to monitor the behaviours of the victim. This scenario represents a scan performed by an insider, who can be a legitimate user of the system doing unauthorised activities. The algorithm produced high true positive rate and low false positive rate, and each experiment could be finished within acceptable time despite the large quantity of data it needs to process.

The DCA was also applied to Botnet detection (Al-Hammadi et al., 2008). Botnets are decentralised and distributed networks of subverted machines, controlled by a central commander, namely "botmaster". A single bot is a malicious piece of program that can transfer victim machines into zombie machines once installed. This work demonstrated the application of the DCA to the detection of a single bot, to assess its performance on this novel problem area. The results indicated that the DCA was able to distinguish the bot from the normal processes on a host machine. The DCA was then applied to a benchmark intrusion detection dataset (Gu et al., 2008), KDD Cup 1999 (Hettich & Bay, 1999). KDD 99 has been widely used and understood, and it is one of the only few labelled datasets that are publicly available in the field of intrusion detection. A preliminary comparison is performed, among the DCA, the real-valued NSA with constant detectors and the C4.5 decision tree algorithm. The results show that the DCA produces reasonably good performance compared to others.

The applications of the DCA indicate the strengths of the algorithm as follows: firstly, the DCA does not require a training phase and the knowledge of normality and anomaly is acquired through basic statistical analysis, so the applications may be less time consuming than other supervised learning algorithms; secondly, the DCA performs linear calculations for its computation, making the system low weight and ideally for intrusion detection tasks. Both strengths make the DCA a suitable candidate for intrusion detection tasks, which mainly require high detection speed.

In summary, the DCA has shown reasonable detection accuracy in the past applications, and it has the advantage of not having a training phase shortening the application process and low weight in computation improving detection speed. As a result, if one needs a system such that a high requirement of detection speed is more crucial, the DCA can be a suitable solution.

**THE DENDRITIC CELL ALGORITHM**

The DCA is a population-based algorithm, designed for tackling anomaly-based detection tasks. It is inspired by functions of natural DCs of the innate immune system, which form part of the body's first line of defence against invaders. DCs have the ability to combine a multitude of molecular information and to interpret this information for the T-cells of the adaptive immune system, to induce appropriate immune responses towards perceived threats. Therefore, DCs can be seen as detectors for different policing sites of the body as well as mediators for inducing a variety of immune responses.

**Algorithm Overview**

Signal and antigen are two types of molecular information processed by natural DCs. Signals are collected by DCs from their local environment and consist of indicators of the health of the monitored tissue. DCs exist in one of three states of maturation to throughout their lifespans. In the initial immature state, DCs are exposed to a combination of signals. Based on the concentration of various signals, DCs can differentiate into either a "fully mature" form to activate the adaptive immune system, or a "semi-mature" form to suppress it. During their immature phase DCs also collect debris in the tissue which are subsequently combined with the environmental signals. Some of the "suspicious" debris collected are termed antigens, and they are proteins originating from potential invading entities. DCs combine the "suspect" antigens with evidence in the form of processed signals to correctly instruct the adaptive immune system to respond, or become tolerant to the antigens in question. For more information regarding the underlying biological mechanisms, please refer to (Greensmith, 2007 and Lutz and Schuler, 2002).
The DCA incorporates the functionality of DCs including data fusion, state differentiation and information temporal correlation. For the remainder of the chapter the term "DC" will refer to the artificial agent based cell, not the natural DCs. For the DCs in the DCA, as per the natural system, there are two types of input data, signal and antigen. Signals are represented as vectors of real-valued numbers, which are measures of the monitored system's status within certain time periods. We term this information as "system context" data. Antigens are categorical values that can be various states of a problem domain or the entities of interest associated with a monitored system. In real world applications, antigens represent what to be classified within a problem domain. For example, they can be process IDs in computer security problems (Al-Hammadi et al., 2008, Greensmith and Aickelin, 2007), a small range of positions and orientations of robots (Oates et al., 2007), the proximity sensors of online robotic systems (mokhtar2009), or the time stamps of records collected in biometric datasets (Gu et al., 2009). Signals represent system context of a host or a measure of network traffic (Al-Hammadi et al., 2008, Greensmith and Aickelin, 2007), measurements derived from various sensors in robotic systems (Oates et al., 2007, Mokhtar et al., 2009), or the biometric data captured from a monitored automobile driver (Gu et al., 2009). Signals are normally pre-categorised as "PAMP", "Danger" or "Safe" in the DCA. The semantics of these signal categories is listed as follows.

- PAMP: a measure that increases in value as the observation of anomalous behaviour. It is a confident indicator of anomaly, which usually presented as signatures of the events that can definitely cause damage to the system.
- Danger: a measure indicates a potential abnormality. The value increases as the confidence of the monitored system being in abnormal status increases accordingly.
- Safe: a measure that increases value in conjunction with observed normal behaviour. This is a confident indicator of normal, predictable or steady-state system behaviour.

Increases in the safe signal value suppress the effects of the PAMP and danger signals within the algorithm, as per what is observed in the natural system. This immunological property has been incorporated within the DCA in the form of predefined weights for each signal category, for the transformation from input signals to output signals. The output signals are used to evaluate the status of the monitored system, to determine the presence of anomalies or misbehaviours.

A relationship of cause-and-effect is believed to exist between signals and antigens, where signals are the explicit effects that potentially result from the implicit cause of antigens. This is achieved if the input data is correctly mapped to the underlying problem domain. The goal of the DCA is to incorporate such a relationship to identify antigens that are responsible for the anomalies reflected by signals. Therefore, the algorithm operates in two steps, firstly it identifies whether anomalies occurred in the past based on the input data; secondly it correlates the identified anomalies with the potential causes, generating an anomaly scene per suspect.

This is achieved by deploying a population of artificial cells, DC objects, which operate in parallel as detectors. Diversity is generated within the DC population through the application of lifespans, which limit the amount of information an individual DC object can process. Different DCs are given different limits for their lifespan, which creates a variable time window effect, with different DC objects processing the signal and antigen data sources during different time periods across the analysed time series (Oates et al., 2008). It is postulated that the combination of signal/antigen temporal correlation and diversity of the DC population are responsible for the detection capability of the DCA.

**Development Pathway**

In this section a brief history is given of the development of the DCA and the numerous versions that have appeared over the past few years. An overview of this process is given in Figure 1.

Figure 1. Development pathway of the DCA.

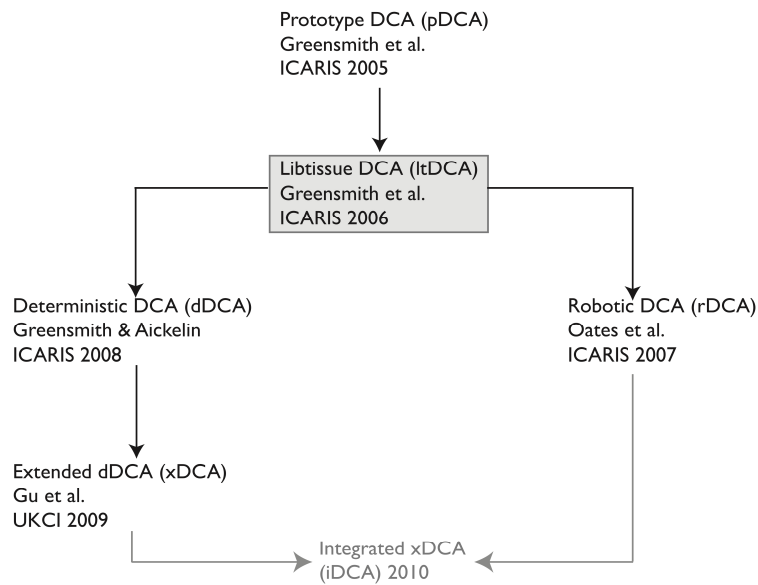

Following the development of the initial abstract model the applicability of the DCA was first demonstrated in a prototype system (pDCA) (Greensmith, 2007). Here the pDCA was applied to a binary classification problem which demonstrated this population based algorithm was capable of performing two-class discrimination on an ordered dataset, using a timestamp as antigen and a combination of features forming three signal categories.

After the encouraging results achieved, the pDCA was further developed into a larger system, combined with the immune-inspired agent based framework libtissue. This version (ltDCA) was stochastic in nature and contained numerous somewhat unpredictable elements including random sampling of incoming antigen, signal decay and randomly assigned migration thresholds. While ltDCA yielded positive results in a number of applications, it contained too many arbitrary and random components, rendering detailed study of its behaviour as an algorithm almost infeasible. Initially the ltDCA was applied to problems in computer and network security including port scan detection and sensor network security.

In parallel to the development of ltDCA, this algorithm steadily increased in popularity especially within the AIS community, being applied to a range of problems. This includes its use as a robotic classifier for physical security. To adapt the ltDCA for purpose, to obtain a reasonable mapping of signal and antigen data, a number of modifications were made, including the segmentation of the output information to provide localisation information for a detected anomaly and the integration of a re-implemented ltDCA with a Brookes style 'subsumption architecture'. This experiment in effect validated the use of the DCA as an anomaly detection technique and transformed its analysis stage into an online process.

Despite its successful application in a number of problem domains, several issues arose in particular with understanding the nature of the "signal and antigen" model of input, and how to interface the categories with given input data. Definitions for the data streams, initially heavily reliant on biological metaphors, were never sufficiently clear and the stochastic nature of ltDCA exacerbated this problem. Theoretical analysis implied that the ltDCA was simply an ensemble linear classifier, which contradicted some of the claims made previously about the manner by which the ltDCA functions.

In response to this, a large amount of randomness was removed from the algorithm and a simplified version developed, named the deterministic DCA (dDCA). This DCA variant is predictable in its use; one can follow a single antigen element throughout the system, and given a priori knowledge of the signals used and the antigen's relative position to all other antigen it is possible to predict its classification. It is shown that the dDCA does not have impaired performance in comparison to ltDCA, and it is possible to calculate crucial algorithmic information such as its real-time capability and run-time complexity values. Still, the dDCA does not avoid the problems that accompany its interface to underlying datasets, nor does it solve the problem of having a non-adaptive analysis module. We propose that ultimately it is necessary to develop an integrated DCA (iDCA) which can automatically extract and select relevant components of underlying data and can adapt to appropriately produce anomaly discrimination in real time and adapt to changes in the underlying dataset. The difference in the dDCA in comparison to a hypothetical iDCA are numerous, therefore we propose to initially develop an intermediate version extended from the dDCA, termed xDCA which is described and used in this paper.

## IMPROVEMENTS TO THE ORIGINAL DCA

Recently a series of improvements to the original DCA have been made, to respond to some of the criticisms. Firstly, a formal definition of the DCA is given to avoid the potential ambiguities of the algorithm. Secondly, an online analysis component based segmentation approaches is introduced to enable periodic and continuous analysis. Thirdly, automated data preprocessing methods based on dimensionality reduction and statistical inference techniques for automated signal selection and categorisation.

### Formalisation of the Algorithm

One criticism of the DCA is the lack of a formal definition, which results in ambiguity for understanding the algorithm and thus leads to incorrect applications and implementations. A formal definition of the DCA should be provided for clearly presenting the algorithm. In this section, we define data structures of the DCA and formalise the algorithm. We concentrate on specifying the entire DC population using quantitative measures at the functional level, and basic set theory and computational functions such as addition, multiplication and recursion are used for clarity.

*Data Structures*

Define $\textbf{Signal} \subseteq \textbf{R}^3$ and $\textbf{Antigen} \subseteq \textbf{N}$ to be the two types of input data. Within a time period ($\textbf{Time} \subseteq \textbf{R}$), the input data can be defined as $S : \textbf{Time} \rightarrow \textbf{Signal} \cup \textbf{Antigen}$, and $S(t)$ is an instance of input data where $t \in \textbf{Time}$. Elements of $\textbf{Signal}$ are corresponding to three signal categories of the DCA as mentioned, which are represented as three-dimensional (3D) vectors and usually normalised into a predefined nonnegative range. Elements of $\textbf{Antigen}$ are categorical values, represented as natural numbers where ordering is ignored.
Define the weight matrix of signal transformation as $W$, a matrix with two rows and three columns, where $w_{ij} \in \textbf{R}$. The weight matrix is used to transform three categories of input signal to two categories output signals. It is usually predefined by users, and kept constant during runtime. The entries are based on empirical results from the underlying immunology. We use the notation $\textbf{List}(A)$ for a list of type $A$. Let $\textbf{Population}$ be a set of DC objects and

$N = |\mathbf{Population}|$ be the population size, a DC object consists of:
- a unique index $i \in \{1, 2, ..., N\}$;
- a lifespan $I : \mathbf{Population} \rightarrow \mathbf{R}$;
- an initial value of its signal profile $K : \mathbf{Population} \rightarrow \mathbf{R}$, which is equal to zero;
- a list **List(Antigen)** for storing its antigen profile.

The signal profile is a measure of processed signal instances, whereas the antigen profile is a measure of sampled antigen instances. Each DC object is able to update its internal data structures by calling relevant operations.

The output of each DC is stored in a list of $\mathbf{N} \times \mathbf{R}$. We use $lst(j)$ to denote the $j$th element of $lst$, and $\pi_1$ and $\pi_2$ for projection functions to get the first and second dimension of a 2-D vector respectively.

*Procedural Operations*

To access the data structures of the DCA, a series of procedural operations are called. These procedural operations describe behaviours of the algorithm at the functional level. They are the most fundamental elements of the algorithm, each of which is executed in one step. The aim of formalising these operations is to present them as simple and clear as possible without losing details.

Let $Append(x, X)$ be a generic function of appending an element $x$ (2-D vector) to a list $X$ (a list of 2-D vectors). The type of $x$ determines the type of $X$, they can be $\mathbf{N} \times \mathbf{R}$ or $\mathbf{R} \times \mathbf{R}$. It is different from "unite" of sets, as duplicates are not eliminated. The function operates by appending the new element to the current list, so it is executed in one step.

At the beginning ($t = 0$), the algorithm initialises all DC objects in **Population** by assigning the initial values of $I(i)$ and $K(i)$, namely "DC initialisation". The value of $I(i)$ depends on the density function used to generate the initial lifespan of each DC object. Both uniform distribution and Gaussian distribution can be applied.

Definition 1 (signal transformation). *The signal transformation function is defined as* $O : \mathbf{Time} \rightarrow \mathbf{R} \times \mathbf{R}$.

$$O(t) = \begin{cases} S(t) \bullet W^\mathrm{T}, & S(t) \in \mathbf{Signal}; \\ \mathbf{Null}, & \text{otherwise,} \end{cases}$$

where $\bullet$ is a multiplication operator for two matrixes. This operation is executed whenever $S(t) \in \mathbf{Signal}$ holds, and it performs a multiplication between a 3-D vector and a transposed $2 \times 3$ matrix to produce another 2-D vector, which consists of two output signals, namely "CSM" and "K". In the case that $S(t) \in \mathbf{Antigen}$ holds, the function returns **Null**.

Definition 2 (lifespan update). *The lifespan update function is defined as* $F : \mathbf{Time} \times \mathbf{Population} \rightarrow \mathbf{R}$.

$$F(t, i) = \begin{cases} I(i), & t = 0; \\ I(i) - \pi_1(O(t)), & F(t-1, i) \leq 0; \\ F(t-1, i) - \pi_1(O(t)), & \text{otherwise.} \end{cases}$$

When $t = 0$, the initial value of $F$ is $I(i)$ which is the initial lifespan of a DC object. It is repeatedly subtracted by *CSM* signal until the termination condition, $F(t-1, i) \leq 0$, is reached, and then it is reset to its initial value $I(i)$.

Definition 3 (signal profile update). *The signal profile update function is defines as* $G : \mathbf{Time} \times \mathbf{Population} \rightarrow \mathbf{R}$.

$$G(t,i) = \begin{cases} K(i), & t = 0; \\ K(i) + \pi_2(O(t)), & F(t-1,i) \leq 0; \\ G(t-1,i) + \pi_2(O(t)), & \text{otherwise.} \end{cases}$$

When $t = 0$, the value of $G$ is $K(i) = 0$ which is the initial signal profile of a DC object. It is repeated added by $k$ signal the termination condition is reached, and the it is reset to its initial value $K(i) = 0$.

Definition 4 (antigen profile update). *The antigen profile update function is defined as* $H : \textbf{Time} \times \textbf{Population} \rightarrow \textbf{List}(\textbf{Antigen})$.

$$H(t,i) = \begin{cases} Append(S(t), H(t-1,i)), & S(t) \in \textbf{Antigen}; \\ \textbf{Null}, & \text{otherwise.} \end{cases}$$

Initially $H$ is empty, and when a new antigen instance arrives, it is sampled by a DC object and stored into its list until the termination condition is reached. The index of the DC object selected is defined as $i \equiv \theta(\bmod N)$, where $\theta$ is the number of antigen instances up to time $t$. This is termed the "sequential sampling" rule.

Definition 5 (output record). *Let* $R(t,i) = \{(r,a) | r = G(t,i) \wedge a \in H(t,i)\}$ *be the output of a DC object, and the output record function is defined as*
$(F(t,i) \leq 0) \wedge (t > 0) \Rightarrow Append(R(t,i), lst)$.

Where $r$ is the signal profile and $a$ is the antigen profile recorded by a DC object. This function is responsible for appending the output of a DC object (a 2-D vector) into the output list *lst* when the termination condition is reached. The processed information in *lst* is used to produce the final detection results in the analysis phase.

Definition 6 (antigen counter). *The antigen counter function for each antigen type* $\alpha \in \textbf{Antigen}$ *is defined as* $C : \textbf{N} \times \textbf{N} \rightarrow \{0,1\}$.

$$C(j,\alpha) = \begin{cases} 1, & \pi_1(lst(j)) = \alpha; \\ 0, & \text{otherwise.} \end{cases}$$

Definition 7 (signal profile abstraction). *The signal profile abstraction function is defined as* $R : \textbf{N} \times \textbf{N} \rightarrow \textbf{R}$.

$$R(j,\alpha) = \begin{cases} \pi_2(lst(j)), & \pi_1(lst(j)) = \alpha; \\ 0, & \text{otherwise.} \end{cases}$$

Function $C$ counts the number of instances of antigen type $\alpha$, and function $R$ calculates the sum of all $k$ values associated with antigen type $\alpha$. These two operations are performed for every single antigen type, and involve scanning *lst* in its entirety.

Definition 8 (anomaly metric calculation). *The anomaly metric calculation function of $K(\alpha)$ is defined as.*

$$\beta = \sum_{j=0}^{n-1} C(j,\alpha) \wedge \gamma = \sum_{j=0}^{n-1} R(j,\alpha)$$

$$\Rightarrow K(\alpha) = \frac{\gamma}{\beta}$$

As $\alpha$ is a known antigen type, we have $\beta \geq 1$. A threshold can be applied for further classification. The value of the threshold depends on the underlying characteristics of the dataset used. If $K(\alpha)$ is greater than the predefined threshold, it is classified as anomalous, other normal.

The procedural operations of the DCA are formally defined in previous section, so by combining them with for or while loops and if statements the algorithm can be presented as the following Pseudocode.

```
foreach DC object do
    DC initilisation;
```

```
        end
while input data do
        if antigen then
                antigen profile update;
        end
        if signal then
                signal transformation;
                foreach DC object do
                        lifespan update;
                        signal profile update;
                        if termination condition then
                                output record;
                        end
                end
        end
end
while ouput list do
        foreach antigen type do
                antigen counter;
                signal profile abstraction;
                anomaly metric cacluation;
        end
end
```

**Online Analysis Component**

An online analysis component is essential for developing an effective intrusion detection system using the DCA. Such a component performs periodic analysis of the processed information presented by DCs, to continuously identify intrusions during detection. An effective and fully functioning intrusion detection system should be able to identify the intrusions as quickly as possible, as accurately as possible, and hence detection speed and detection accuracy are two major indicators of performance. By detection speed here we meant the time an intrusion detection system takes to identify intrusions or anomalies during detection. Most techniques can produce reasonable detection accuracy, if sufficient time is given. But in most case detection speed is more significant for assessing the performance of an intrusion detection system. If the system fails to identify the intrusions in time, no further responses against the intrusions can be made. This leads to the eventual success of attacks, which is a fatal failure of an intrusion detection system. In other words, if the intrusions are identified too late, even with 100% detection accuracy, it all becomes meaningless in terms of system defence. As a result, we propose integrating online analysis with the DCA, to improve detection speed without compromising detection accuracy.

If online analysis is to be performed during detection, one issue needs resolved, namely, when to perform analysis. This could be solved by introducing segmentation to the DCA. It is different from the moving time windows method described in (Gu et al., 2008), which is used in the pre-processing stage to smooth noisy input signals, as segmentation is performed in the post-processing stage for the purpose of periodic and continuous analysis. As the processed information is presented by matured DCs over time, a sequence of processed information is being generated during detection. Segmentation involves partitioning this sequence into relative smaller segments, in terms of the number of data items or time. All the generated segments have the same size, and the analysis is performed within each individual segment. Therefore, in each segment, one set of detection result ($K(\alpha)$ per antigen type) is generated, in which intrusions appeared within the duration of this segment can be identified.

First of all, segmentation can produce multiple sets of results, rather than one set of results produced by non-segmentation system. This enables the system to perform analysis online not offline, as all segments are processed during detection. In addition, segmentation distributes the analysis process into multiple steps, instead of performing at once. This can reduce the computation power and time required for the analysis process, so segmentation can effectively enhance detection speed. Moreover, as the processed information is presented by matured DCs at different time points over the duration, analysing the sequence of processed information at once ignores the temporal difference of each piece of processed information. As a result, the same antigen type that causes malicious activities at one point but does nothing at another point may be classified as normal rather than an intrusion. This can be avoided by applying segmentation, as it features periodic analysis that can cope with the inherited time differences. Therefore, the system can effectively discriminate malicious activities from normal ones, and hence the detection accuracy can also be improved.

The most important and in fact the only factor of segmentation is the segment size. It determines how soon the intrusions can be identified. The smaller the segment size, the sooner the intrusions can be identified, and vice versa. Moreover, the segment size may also influence the sensitivity of the final results. If the segment size is too large, the results can lose the sensitivity and thus the system loses the ability to identify true positives. However, if the segment size is too small, the results may be too sensitive, and the system can generate false positives.

In (Gu et al., 2009), the authors have shown that applying segmentation to the DCA makes significant differences to the results. In some cases, the system with segmentation may even produce better performance in terms of detection accuracy. In addition, segmentation enables the system to perform periodic analysis on the processed information presented by the DCs. As a result it can effectively improve detection speed without compromising detection accuracy. Therefore, segmentation is applicable to the DCA. Even though segmentation is not immune inspired, it can still make contribution to the field of AIS, as it can improve the system performance of the DCA. As a result, more effective intrusion detection systems can be developed, by integrating segmentation with the DCA. This method is also applicable to other second generation AIS algorithms.

**Automated Data Preprocessing**

The DCA encounters issues that accompany its interface to underlying datasets, as well as have a non-adaptive analysis module. Formal definition and theoretical analysis have shown that the algorithm itself is rather simple, and segmentation enables the DCA to efficiently and effectively process large datasets in terms of data size. However, data size is not the only concern when handling complex datasets, high dimensionality is often a bigger problem. Complexity occurs at the data pre-processing stage of the DCA when dimensionality reduction is required. Previously, the data pre-processing of the DCA is performed manually based on users' expert knowledge of a given problem domain, which is time consuming and sometime difficult to achieve. As a result, it is necessary to automate the data pre-processing stage, which extracts and selects relevant features, and adapts the algorithm to characteristics of the underlying data.

Originally the DCA does not rely on training data to define which of the input signals are potentially "dangerous". But expert knowledge of the problem domain is required to generate proper input data to the DCA during manual data pre-processing. It involves two steps, namely "antigen representation" and "signal selection and categorisation" for generating two types of input to the DCA. Since antigen representation involves identifying the objects to classify as normal or anomalous in the problem, it is usually easy to accomplish and thus does not require automation. The rest of this section is focused on the automation of signal selection and categorisation in the DCA.

Signal selection and categorisation involve firstly selecting or extracting the most interesting features from the original feature set of a given dataset, and then categorising these features into one of three signal categories of the DCA. Manual methods are problematic, as whenever the problem is changed, the whole process needs to be redone. Even for the same dataset, if cross-validation is used, the process needs to be performed for every single subset generated. In addition, it is not always possible to acquire adequate expert knowledge for a dataset to effectively perform signal selection and categorisation for the DCA. In order to automate signal selection and categorisation, techniques of dimensionality reduction and statistical inference are used, and an unsupervised "training phase" is introduced to obtain the underlying knowledge regarding normality and anomaly of a problem.

Let $X = \{\mathbf{x}_1, \mathbf{x}_2, ..., \mathbf{x}_m\}$ the original feature set of a problem. Dimensionality reduction involves generating a new feature space $Y = \{\mathbf{y}_1, \mathbf{y}_2, ..., \mathbf{y}_d\}$ ($d < m$), which is supposed to be representative of the problem with minimum redundancy. Feature selection involves selecting a best subset of input features. Feature extraction creates new features based on transformations or combinations of the original features. The choice between feature selection and extraction depends on the problem domain and classifier. The possible techniques are listed as follows.

- Correlation Coefficient (with class labels) - feature selection
- Information Gain (with class labels) - feature selection
- Principal Component Analysis (PCA, without class labels) - feature extraction

Preliminary work has been carried out in Gu et al. (2009), in which the authors showed that it was possible to integrate PCA with the DCA for the purpose of automated data preprocessing. The PCA facilitateed the reduction of data dimension of the raw data, to select proper features as the candidates of the input of the DCA. It was also used for the ranking of attributes based on the variability, which is mapped to the ranking of signal categories of the DCA for signal categorisation. In this way, the data preprocessing of the DCA is performed by simply using PCA and basic Min-Max normalisation, without the requirement of any expert knowledge of the problem domain. The results suggested that the integrated system of PCA and the DCA was successful in terms of anomaly detection, as the system can produce relatively high true positive rates and low false positive rates. As a result, the application of PCA to the DCA makes it possible to automatically categorise input data into user-defined signal categories, while still generating useful and accurate classification results.

More recently, automated data preprocessing methods based on feature selection/extraction and statistical inference techniques are developed. The idea is to use dimensionality reduction techniques to select the *most interesting* features from the original feature set as the candidates of input signals to the DCA, then perform greed search to find the best combination of these candidates corresponding to the signals categories where mean square errors (MSE) (Garthwaite et al., 2006) is employed as a performance measure. The new system is tested by a number of datasets, as well as compared with other existing machine learning techniques. The testing datasets include: the UCI Wisconsin Breast Cancer dataset (Blake et al., 1998) consists of 700 data instances and each data item has 10 features; the KDD 99 dataset derived from the DAPRA 98 Lincoln Lab data set for applying data mining techniques to the area of intrusion detection, it consists of about 5 million data instances, each of which has 42 features. Two versions of the KDD 99 dataset are used: a 10% subset whose data items are randomly and proportionally selected from the whole dataset; and the whole dataset. As these two datasets are much more complicated, 10-fold cross validation is performed to generate 10 subsets. For the purpose of baseline comparison, other existing techniques, including K-Nearest-Neighbour (KNN), decision trees, and Support Vector Machine (SVM) algorithms are used.

Table 1 shows the true positive rate and false positive rate of all methods applied to the breast cancer dataset. As expected, manual method produces high true positive rate and low false positive rate; PCA based method produces better result since its true positive rate is higher and false positive rate is zero; information gain based method has significantly lower true positive rate but comparable false positive rate; correlation based method has comparable true positive rate but significantly higher false positive rate.

Table 1. Results of all methods on the breast cancer dataset.

|  | True Positive Rate | False Positive Rate |
|---|---|---|
| **Manual** | 0.963 | 0.033 |
| **PCA** | 0.985 | 0.000 |
| **Information Gain** | 0.789 | 0.008 |
| **Correlation** | 0.930 | 0.213 |

As shown in Figure 2 and Figure 3, the manual method and other existing techniques produce good performance in terms of both true positive rate and false positive rate. Whereas, automated methods produce the results that vary from one subset to another. The PCA based method produces high true positive rate and low false positive rate for most subsets, except one where the true positive rate is significantly lower. The information gain based method produces true positive rates within a relatively large range but still all above the boundary of random classifier (50%), and it keeps the false positive rates at a low level. The correlation based method produces high true positive rates across, but simultaneously high false positive rates across all the subsets. The results show that it is possible to automate the data preprocessing of the DCA to produce useful. However, as this is work in progress, more investigation should be performed.

Figure 2. Boxplot of true positive rates of all methods for KDD 10% dataset.

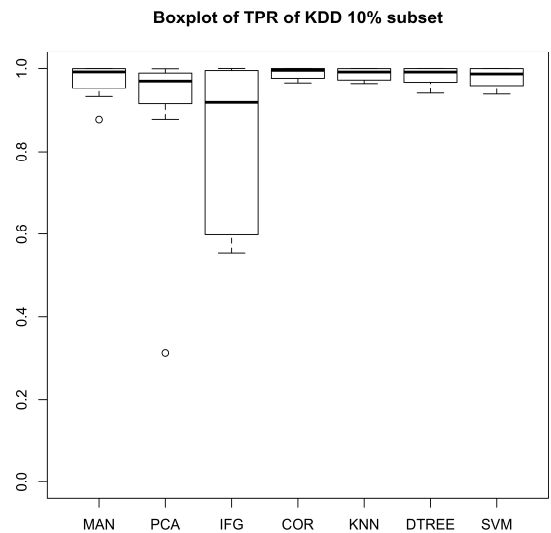

Figure 3. Boxplot of false positive rates of all methods for KDD 10% dataset.

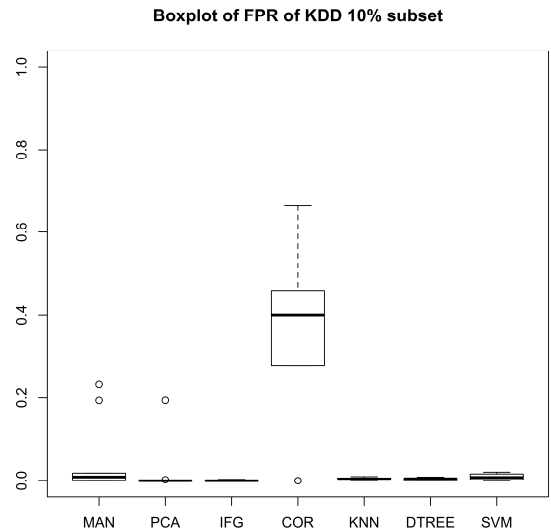

CONCLUSION AND FUTURE DIRECTIONS

In this book chapter, we reviewed some of the most popular AIS algorithms for intrusion detection problems, as well as demonstrating that the DCA is a suitable solution due to its unsupervised learning paradigm and low weight in computation. In order to present the DCA as simply and clearly as possible without losing detail, we used set theory and basic functions to formalise the algorithm. In addition, we demonstrated recent work for introducing an online analysis component to the DCA through segmentation, and the results indicate that the approach is promising especially for intrusion detection problems. Finally, ongoing work of automated data preprocessing of the DCA was described. With the automated data preprocessing, applications of the DCA are not dependent on the expert knowledge of the problem, and they become much less time consuming since manual procedures are replaced. Much of the work regarding the further development of the DCA is still 'work in progress'. For example, only static segment sizes applied and tested for the online analysis component of the DCA, more adaptive mechanisms where segment size varies according the situations encountered during detection should be investigated. Additionally, only preliminary work has been performed for the automated data preprocessing of the DCA, more techniques of feature selection and feature extraction and more appropriate mechanisms for signal categorisation should be evaluated.